\documentclass[twocolumn]{aastex62}

\usepackage[normalem]{ulem}
\useunder{\uline}{\ul}{}
\usepackage{hyperref}
\usepackage{lineno}

\usepackage{color}
\usepackage{amsmath,amssymb}

\hypersetup{linkcolor=cyan,citecolor=magenta,filecolor=yellow,urlcolor=blue}

\shorttitle{GRB 190829A}
\shortauthors{Yu Wang et al.}

\begin{document}

\title{GRB 190829A - A Showcase of Binary Late Evolution}

\author{Yu~Wang}
\affiliation{ICRA, Dip. di Fisica, Universit\`a  di Roma ``La Sapienza'', Piazzale Aldo Moro 5, I-00185 Roma, Italy}
\affiliation{ICRANet, Piazza della Repubblica 10, I-65122 Pescara, Italy} 
\affiliation{INAF -- Osservatorio Astronomico d'Abruzzo,Via M. Maggini snc, I-64100, Teramo, Italy. rahim.moradi@inaf.it}

\author{J.~A.~Rueda}
\affiliation{ICRA, Dip. di Fisica, Universit\`a  di Roma ``La Sapienza'', Piazzale Aldo Moro 5, I-00185 Roma, Italy. yu.wang@uniroma1.it}
\affiliation{ICRANet, Piazza della Repubblica 10, I-65122 Pescara, Italy}
\affiliation{ICRANet-Ferrara, Dip. di Fisica e Scienze della Terra, Universit\`a degli Studi di Ferrara, Via Saragat 1, I--44122 Ferrara, Italy}
\affiliation{Dip. di Fisica e Scienze della Terra, Universit\`a degli Studi di Ferrara, Via Saragat 1, I--44122 Ferrara, Italy}
\affiliation{INAF, Istituto di Astrofisica e Planetologia Spaziali, Via Fosso del Cavaliere 100, 00133 Rome, Italy}

\author{R.~Ruffini}
\affiliation{ICRA, Dip. di Fisica, Universit\`a  di Roma ``La Sapienza'', Piazzale Aldo Moro 5, I-00185 Roma, Italy}
\affiliation{ICRANet, Piazza della Repubblica 10, I-65122 Pescara, Italy}
\affiliation{INAF,Viale del Parco Mellini 84, 00136 Rome, Italy}

\author{R.~Moradi}
\affiliation{ICRA, Dip. di Fisica, Universit\`a  di Roma ``La Sapienza'', Piazzale Aldo Moro 5, I-00185 Roma, Italy. yu.wang@uniroma1.it}
\affiliation{ICRANet, Piazza della Repubblica 10, I-65122 Pescara, Italy} 
\affiliation{INAF -- Osservatorio Astronomico d'Abruzzo,Via M. Maggini snc, I-64100, Teramo, Italy}

\author{Liang~Li}
\affiliation{ICRA, Dip. di Fisica, Universit\`a  di Roma ``La Sapienza'', Piazzale Aldo Moro 5, I-00185 Roma, Italy}
\affiliation{ICRANet, Piazza della Repubblica 10, I-65122 Pescara, Italy} 
\affiliation{INAF -- Osservatorio Astronomico d'Abruzzo,Via M. Maggini snc, I-64100, Teramo, Italy}

\author{Y.~Aimuratov}
\affiliation{ICRA, Dip. di Fisica, Universit\`a  di Roma ``La Sapienza'', Piazzale Aldo Moro 5, I-00185 Roma, Italy}
\affiliation{ICRANet, Piazza della Repubblica 10, I-65122 Pescara, Italy} 
\affiliation{Fesenkov Astrophysical Institute, Observatory 23, 050020 Almaty, Kazakhstan}
\affiliation{al-Farabi Kazakh National University, al-Farabi avenue 71, 050040 Almaty, Kazakhstan}

\author{F.~Rastegarnia}
\affiliation{ICRANet, Piazza della Repubblica 10, I-65122 Pescara, Italy}
\affiliation{Dip. di Fisica e Scienze della Terra, Universit\`a degli Studi di Ferrara, Via Saragat 1, I--44122 Ferrara, Italy}
\affiliation{Department of Physics,Faculty of Physics and Chemistry, Alzahra University,Tehran, Iran}

\author{S.~Eslamzadeh}
\affiliation{ICRANet, Piazza della Repubblica 10, I-65122 Pescara, Italy}
\affiliation{Dip. di Fisica e Scienze della Terra, Universit\`a degli Studi di Ferrara, Via Saragat 1, I--44122 Ferrara, Italy}
\affiliation{Department of Theoretical Physics, Faculty of Basic Sciences, University of Mazandaran, P. O. Box 47416-95447, Babolsar, Iran}

\author{N.~Sahakyan}
\affiliation{ICRANet-Armenia, Marshall Baghramian Avenue 24a, Yerevan 0019, Republic of Armenia}

\author{Yunlong~Zheng }
\affiliation{Department of Astronomy, School of Physical Sciences, University of Science and Technology of China, Hefei, Anhui 230026, China}
\affiliation{CAS Key Laboratory for Researches in Galaxies and Cosmology, University of Science and Technology of China, Hefei, Anhui 230026, China}
\affiliation{School of Astronomy and Space Science, University of Science and Technology of China, Hefei, Anhui 230026, China}
\affiliation{ICRANet, Piazza della Repubblica 10, I-65122 Pescara, Italy}

\email{yu.wang@icranet.org, jorge.rueda@icra.it, ruffini@icra.it,\\ rahim.moradi@icranet.org, liang.li@icranet.org} 

\begin{abstract}
GRB 190829A is the fourth closest gamma-ray burst (GRB) to date ($z=0.0785$). Owing to its wide range of radio, optical, X-ray, and the very-high-energy (VHE) observations by H.E.S.S., it has become an essential new source examined by various models with complementary approaches. We here show in GRB 190829A the double-prompt pulses and the three-multiwavelength afterglows are consistent with the type II binary-driven hypernova (BdHN II) model. The progenitor is a binary composed of a carbon-oxygen (CO) star and a neutron star (NS) companion. The gravitational collapse of the iron core of the CO star produces a supernova (SN) explosion and leaves behind a new neutron star ($\nu$NS) at its center. The accretion of the SN ejecta onto the NS companion and onto the $\nu$NS via matter fallback spins up the NSs and produces the double-peak prompt emission. The synchrotron emission from the expanding SN ejecta with the energy injection from the rapidly spinning $\nu$NS and its subsequently spin-down leads to the afterglow in the radio, optical, and X-ray. We model the sequence of physical and related radiation processes in BdHNe and focus on individuating the binary properties that play the relevant roles.
\end{abstract}


\section{Introduction} 
\label{sec:introduction}

As one of the closest GRBs \citep{2019GCN.25552....1D,2019GCN.25563....1H,2021A&A...646A..50H}, GRB 190829A underwent one of the most extensive observational campaigns, including but not limited to the Fermi satellite \citep{2019GCN.25551....1F}, the Neil Gehrels Swift Observatory \citep{2019GCN.25657....1P}, the High Energy Stereoscopic System (H.E.S.S.) \citep{abdalla2021revealing}, the Gran Telescopio Canarias (GTC) \citep{2019GCN.25677....1D} and the Gamma-Ray Burst Optical/Near-Infrared Detector (GROND) \citep{2019GCN.25651....1B}. GRB 190829A has become a key source to test details of alternative GRB models. The conventional concept of GRBs postulates that when the core of a single massive star collapses, a relativistic jet-like outflow forms and propagate. The internal shock in the outflow produces prompt emissions. Then the outflow interacts with the interstellar medium generating the afterglow via synchrotron process, as well as the very-high-energy (VHE) emission via synchrotron self-Compton process \citep{Meszaros2002,2004RvMP...76.1143P,zhang2018physics,2019Natur.575..464A,2019Natur.575..455M,2019Natur.575..448Z,abdalla2021revealing}. Here, we present an alternative approach: the progenitor is a binary system composed of a carbon-oxygen core (CO$_{\rm core}$) and a companion NS in a tight orbit with orbital period of a few minutes. The iron core of the CO$_{\rm core}$ collapses and generates a supernova (SN) at the end of its thermonuclear evolution, a new $\nu$NS is left at the SN center. The accretion of SN ejecta onto the companion NS and the fallback accretion onto the $\nu$NS contribute to the energy of prompt emission and spins up the $\nu$NS. The rotational energy from the $\nu$NS spin-down powers the afterglow of synchrotron emission \citep{2012ApJ...758L...7R,2014ApJ...793L..36F,2019ApJ...871...14B,2021MNRAS.504.5301R,2021IJMPD..3030007R}. The observed optical SN \citep{2019GCN.25657....1P, 2019GCN.25677....1D} completes this alternative self-consistent approach. 

Specifically, \citet{abdalla2021revealing} presented the H.E.S.S. observations of VHE photons of hundreds of GeV lasting $10^5$~s. The VHE exhibits a luminosity decaying index and a spectral shape similar to the ones of the X-ray afterglow emission. The standard forward shock model was there applied to the afterglow and showed its difficulties in explaining these observations. \citet{2020MNRAS.496.3326R} showed the radio observations could be explained within the synchrotron forward shock model. \citet{2021A&A...646A..50H} presented the optical observations, analyzed the multiwavelength data, and compared this burst with GRB 180728A. \citet{2021ApJ...918...12F} modelled the optical and X-ray observations in the afterglow by the synchrotron forward shock model, and the VHE observations by the synchrotron self-Compton scattering. \citet{2021ApJ...920...55Z} interpreted the VHE observation by the external inverse-Compton scenario with the seed photons coming from the prompt emission pulses. \citet{2020ApJ...898...42C} analyzed various episodes of this burst and concluded that the shockwave breakout model could not explain the entire burst. \citet{2021MNRAS.504.5647S} proposed this GRB was viewed from an off-axis angle in an attempt to solve the dilemma of the VHE photons produced in a low-luminosity GRB.  \citet{2021ApJ...917...95Z} proposed that the interaction of the hard X-ray photons in the first prompt pulse with the dusty medium produces the second prompt pulse and an electron-positron rich medium in which the synchrotron self-Compton produces the VHE. \citet{2021arXiv211114861D} focused on the early afterglow, their multiwavelength studies purported the existence of both forward and reverse shocks.

The above articles present detailed observations including radio, optical, X-ray and VHE, and give a variety of interpretations of the different emission episodes: they all generally assume a single progenitor and ultra-relativistic shock waves. In this article, we start by focusing on the nature of the binary progenitor, and far from describing a single leading ultra-relativistic process, we emphasize the existence of a number of episodes with different emission processes, which we examine in their rest-frame. We do not evidence any ultra-relativistic emission, on the contrary, we evidence: 1) the special role of two early pulses observed by Fermi and Swift relates to the progenitor of binary components; 2) the crucial role of the synchrotron electromagnetic radiation from the mild-relativistic expanding SN ejecta for describing the afterglow composed of radio, optical, and X-ray emissions; 3) we finally address the appearance of the SN, of which the optical emission is brighter than the one from synchrotron. Therefore, in our approach, we model the sequence of physical and related radiation processes, and focus on individuating the binary properties that play the relevant roles.

The binary model was proposed in 2012 \citep{2012ApJ...758L...7R} and has been in development for one decade. The physical picture and modelling of the SN ejecta accretion onto the NS companion have been gradually extended to include the needed physics that allows to study a wide range of binary parameters, based on detailed analysis of multiple well-observed GRBs and the statistical analysis of different GRB components \citep{1999A&A...350..334R, 2000A&A...359..855R,  2010PhR...487....1R,2015ApJ...798...10R, 2015ARep...59..667W, 2018ApJ...869..101R, 2018ApJ...852...53R, 2018MmSAI..89..293W, 2018ApJ...869..151R, 2019ApJ...874...39W, 2019ApJ...886...82R, 2020ApJ...893..148R, 2020EPJC...80..300R, 2021A&A...649A..75M, 2021MNRAS.504.5301R}. The numerical simulations of the occurring physical processes has been upgraded from one-dimension \citep{2014ApJ...793L..36F} to two-dimensions \citep{2015ApJ...812..100B}, to three-dimensions \citep{2016ApJ...833..107B,2019ApJ...871...14B}. The latest simulations \citep{2019ApJ...871...14B} implemented a smoothed-particle-hydrodynamics (SPH) method, and examined a large selection of initial conditions and the outcomes of the binary system after the SN explosion. \citet{2019Univ....5..110R} and \citet{2021IJMPD..3030007R} have reviewed the entire development process. In this article, we have the scenario, namely BdHN II, that the NS does not accrete enough matter to reach the critical mass for black hole (BH) formation, it remains stable as a more massive NS (MNS).

Unlike the traditional fireball model, the BdHN model considers a central engine arises in the final evolutionary stage of the CO$_{\rm core}$ in the presence of a binary companion. A SN explosion occurs, it triggers the GRB emission and generates a $\nu$NS. Therefore, in addition to the physical processes of single-star collapse models, we need to consider not only the binary interactions, but also the appearance of the $\nu$NS. The most influential interactions are the accretion of SN ejecta onto the NS companion and the fallback accretion onto the $\nu$NS spins it up. The afterglow is produced by the mildly-relativistic expanding SN ejecta which contains a large number of electrons accelerated by the kinetic energy of the SN and the energy injection from the rapidly spining $\nu$NS and its subsequently spindown. In this article, we will model the afterglow of GRB 190829A following the above picture. An additional advantage is that it naturally accounts for the observed association of GRBs with type Ic SNe \citep{2012ApJ...758L...7R} and indicates the peak luminosity of the optical SN emission well above the synchrotron optical emission.

This article is structured as follows. In section \ref{sec:picture}, we present the physical picture and the emission episodes that our model predicts. In section \ref{sec:observation}, we introduce the observational data. In section \ref{sec:prompt}, we analyse the prompt emission and explain the prompt pulses by the SN explosion and SN ejecta accretion on the companion NS and $\nu$NS. In section \ref{sec:synch}, we move to the analyse the afterglow, and model the radio, optical and X-ray emissions by the synchrotron emission from the SN ejecta. In section \ref{sec:conclusion}, we present the conclusions of the article.

\section{Physical Picture and Expectation}
\label{sec:picture}

\begin{figure}
\centering
\includegraphics[width=1\hsize,clip]{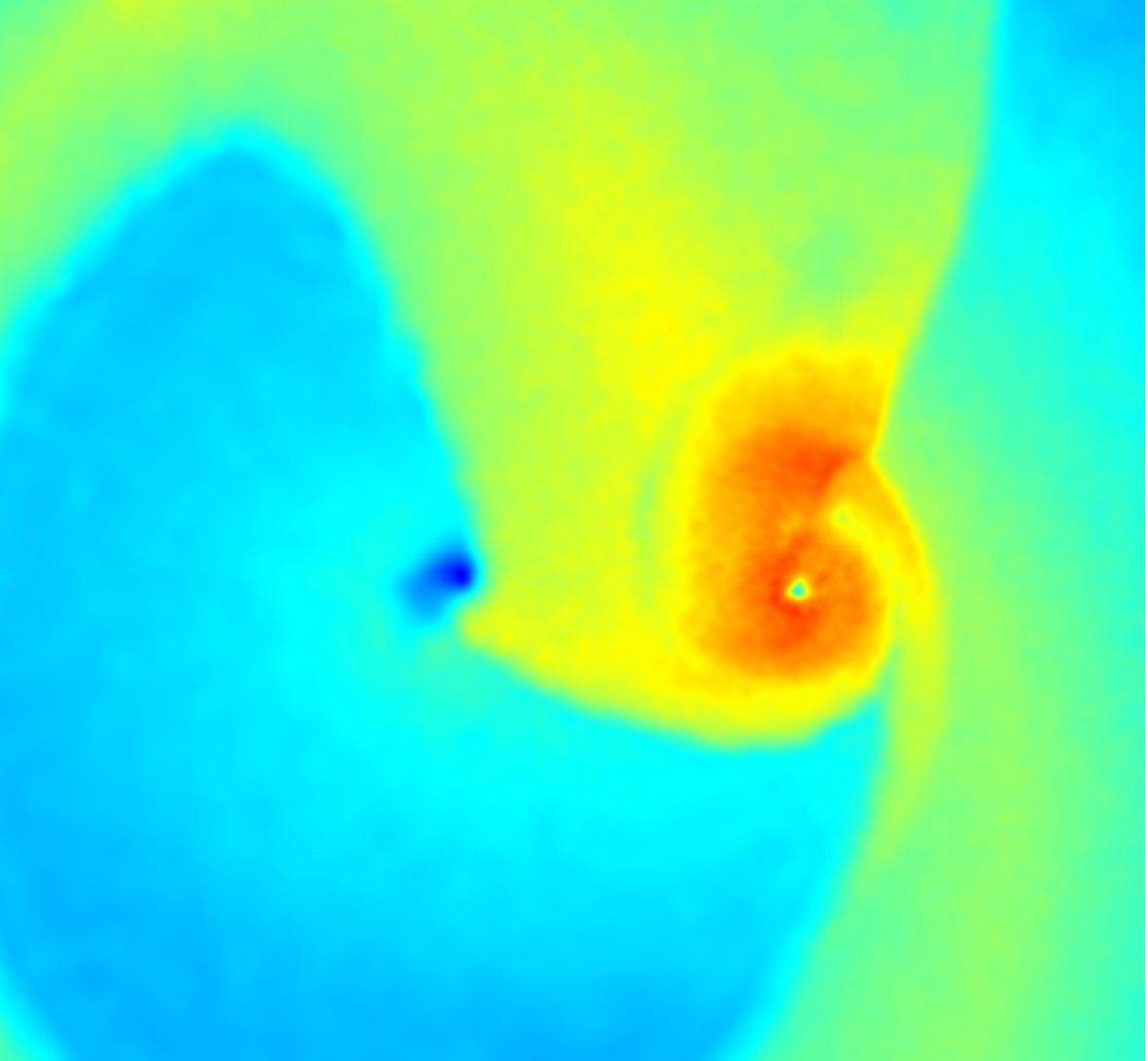}
\caption{Ongoing accretion process onto the $\nu$NS and the NS companion simulated in \citet{2019ApJ...871...14B}. The $\nu$NS is located at the center of the dark blue spot, and is accreting the surrounding material. The SN ejecta are also being accreted by the NS companion, which is located at the center of the green spot. We also notice that the expansion of SN ejecta is distorted by the companion NS and a part of the SN ejecta is flowing back to the $\nu$NS. This process creates a unique feature of BdHNe: the fallback accretion onto the $\nu$NS is enhanced creating a second peak of accretion at about an orbital-period time after the SN explosion \citep[see, e.g., Fig. 5 in ][for more details]{2019ApJ...871...14B}.}
\label{fig:accretion}
\end{figure}

\begin{figure*}
\centering
\includegraphics[width=0.75\hsize,clip]{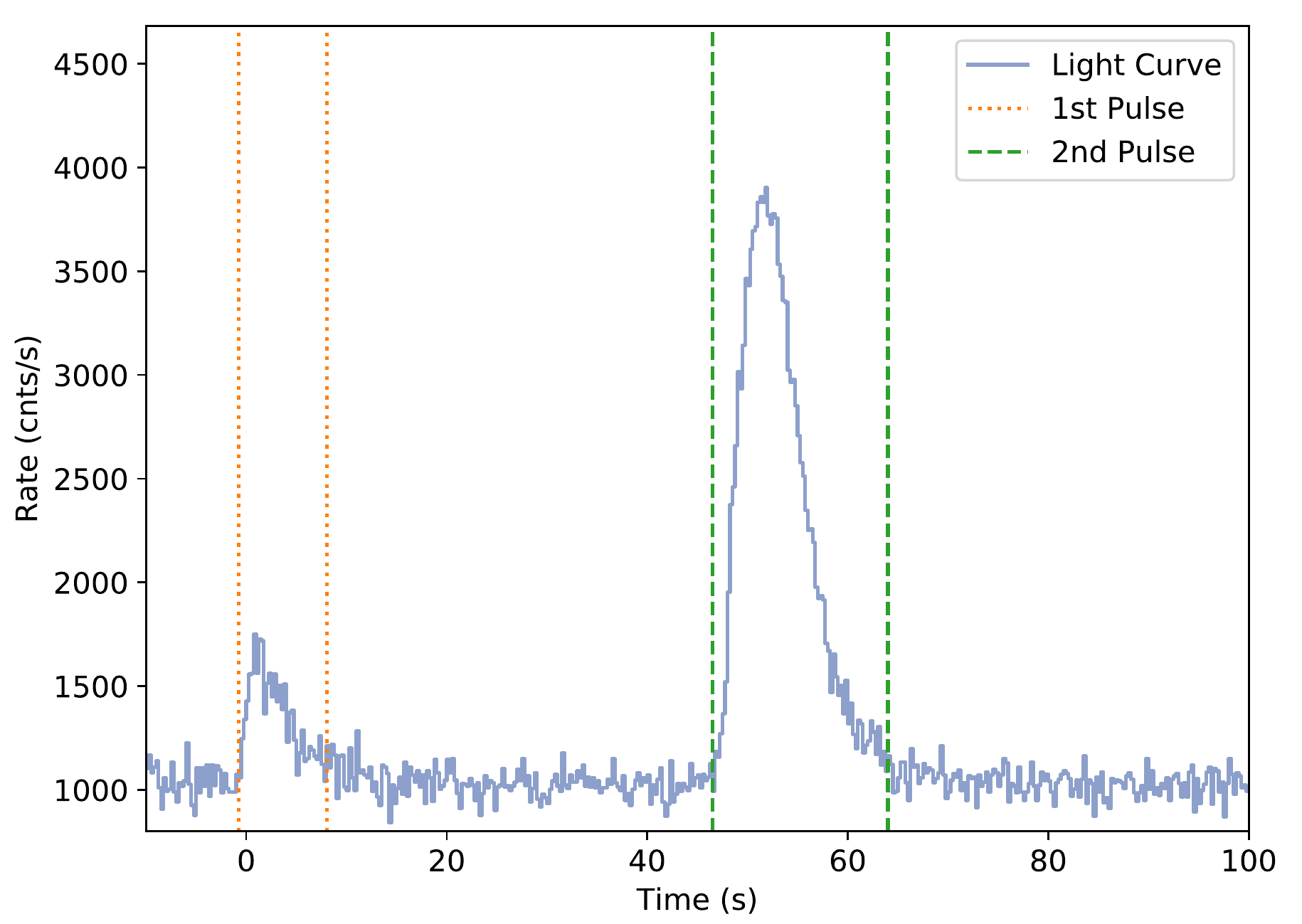}
\caption{The count rate of GRB 190829A prompt emission from the raw data of \textit{Fermi}-GBM: The first pulse is from $-0.75$~s to $8.05$~s, indicated by the orange dotted line, and the second pulse is from  $46.50$~s to $64.00$~s, indicated by the green dashed line.}
\label{fig:prompt_190829A}
\end{figure*}

\begin{figure*}[ht]
\centering
\includegraphics[width=0.85\hsize,clip]{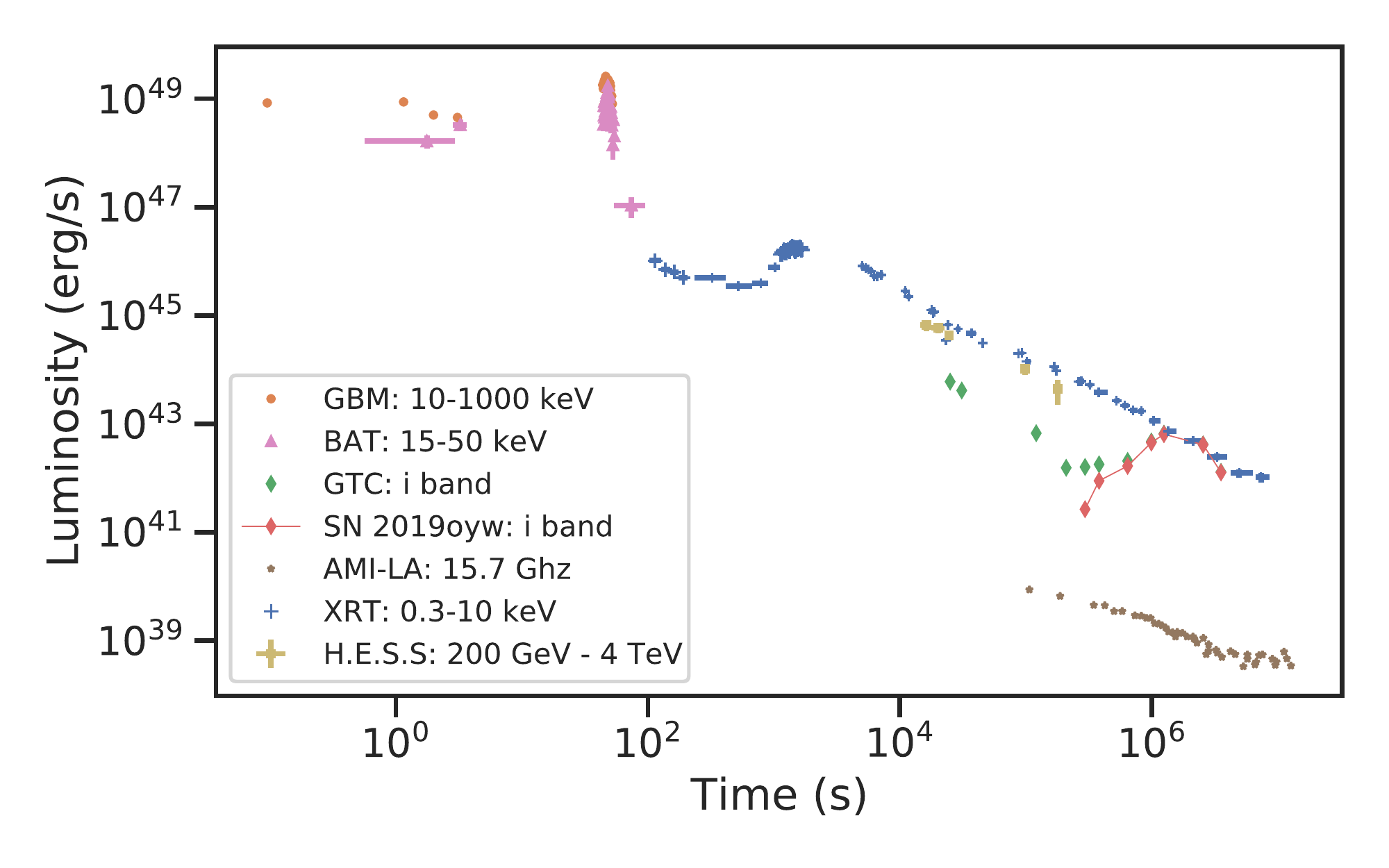}
\caption{Luminosity of GRB 190829A including the data from H.E.S.S (yellow) for TeV, \textit{Fermi}-GBM (orange dots), \textit{Swift}-BAT (purple triangles) for the prompt emission of hard X-ray and gamma-ray, \textit{Swift}-XRT (blue crosses) for the soft X-ray (absorbed), GTC (green diamonds) for the optical i band, from which the SN 2019yw is extracted (red diamonds), the optical signal of SN over-shots the optical emission from the synchrotron, and AMI-LA (brown stars) for the radio observation.}
\label{fig:luminosity_190829A}
\end{figure*}

\begin{figure*}
\centering
\includegraphics[width=0.75\hsize,clip]{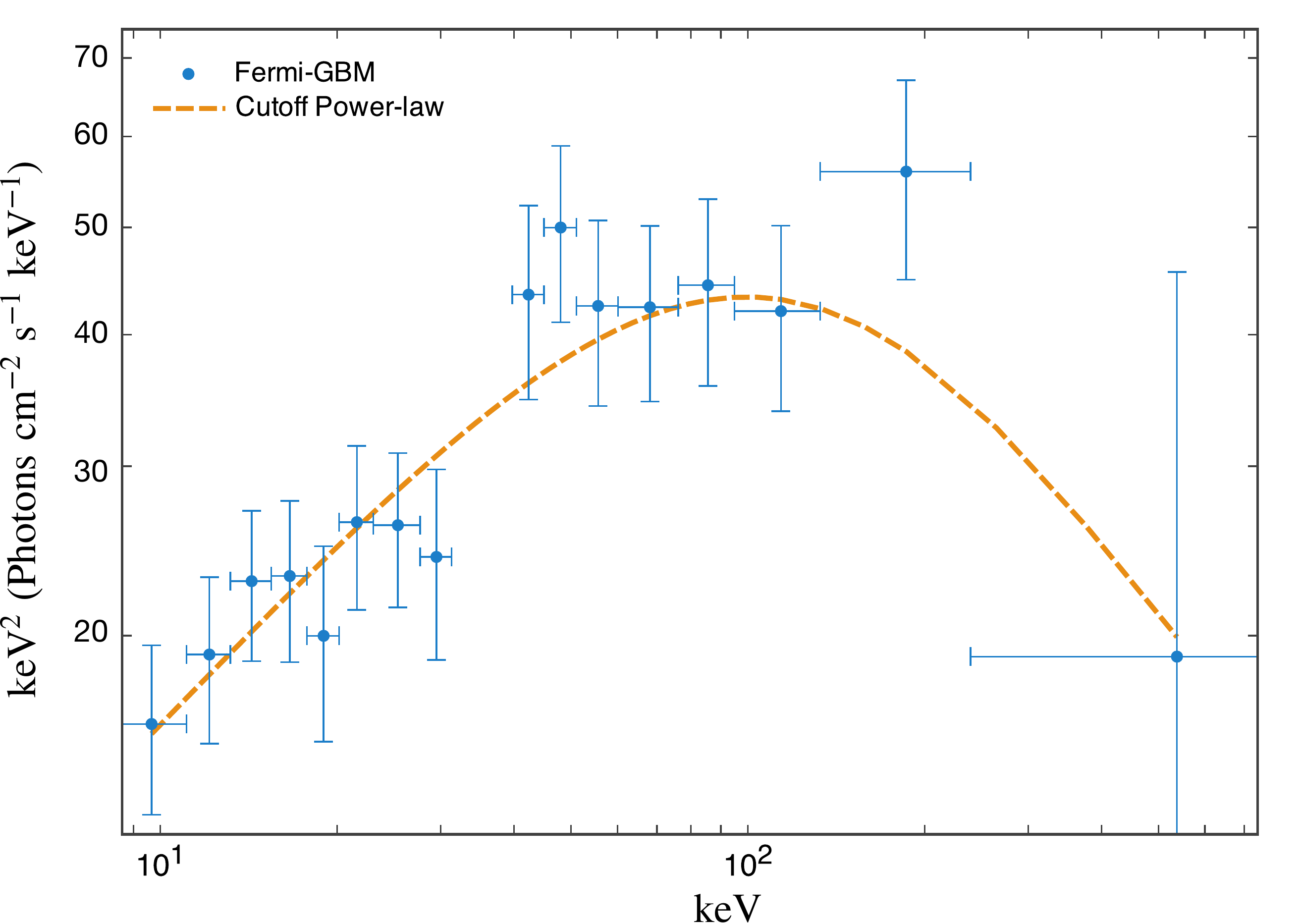}
\includegraphics[width=0.75\hsize,clip]{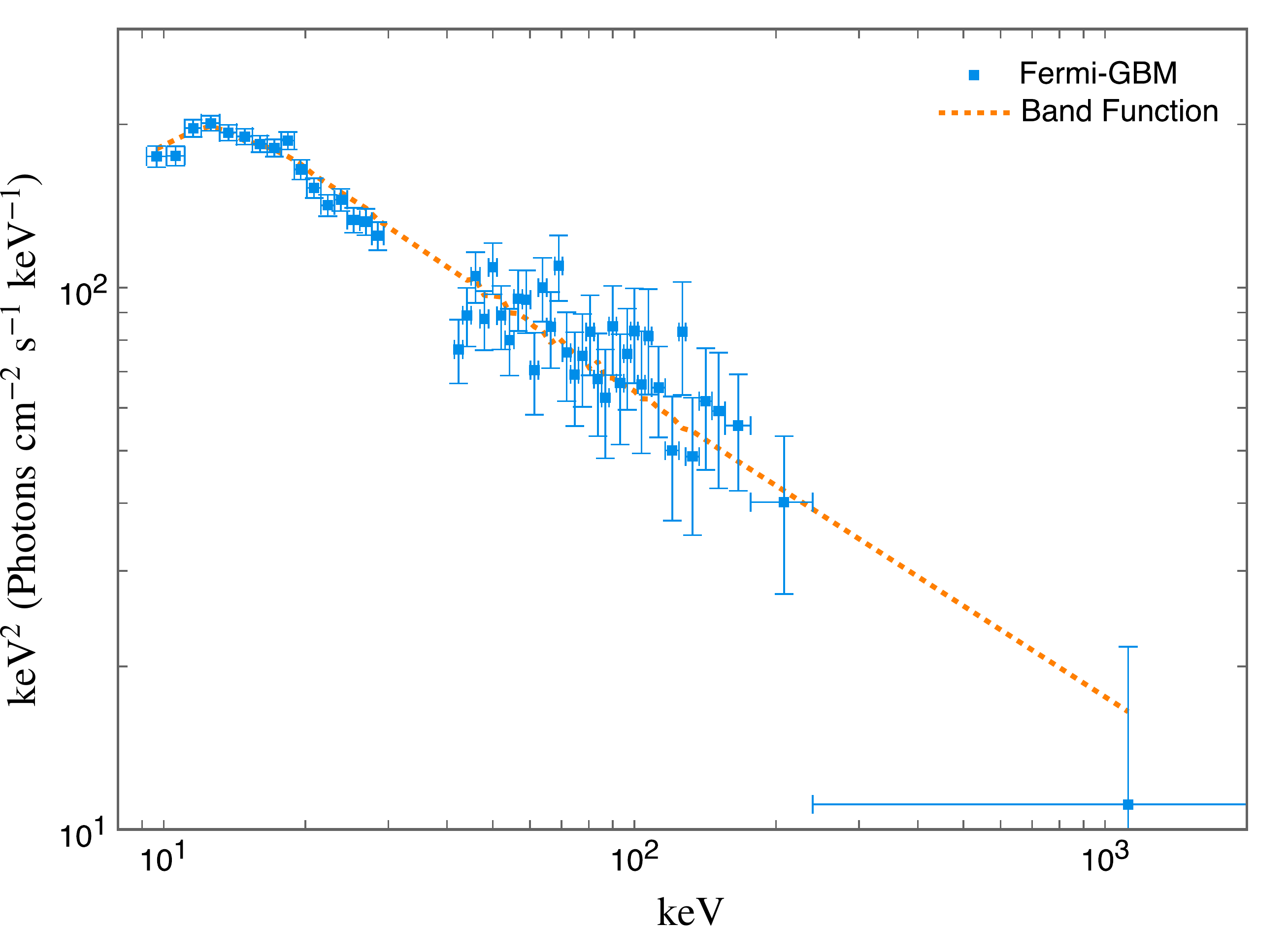}
\caption{\textbf{Top}: Spectrum of the first pulse observed by Fermi-GBM. The blue points are the data and the orange curve indicates the fitting by a  cutoff power-law with power-law index $\alpha = -1.45$ and the peak energy $E_p = 144.28$~keV. \textbf{Bottom}: Spectrum of the second pulse. Blue points are the data and the orange curve indicates a band function fitting, with the low-energy index $\alpha = 0.50$, the high-energy index $\beta = -2.53$, and the peak energy $E_p = 13.58$~keV.}
\label{fig:prompt_spectrum}
\end{figure*}

As recalled in the introduction, we consider a binary system composed of a CO$_{\rm core}$ and a NS with an orbital period of a few tens of minutes \citep{2021MNRAS.504.5301R}.
At a given time, the CO$_{\rm core}$ collapses, forms a newborn NS ($\nu$NS) at its centre, and induces a SN explosion. Most of the SN energy ($\sim 10^{53}$~erg) is deposited in the neutrino, about a few per cent of energy goes to the kinetic energy of SN ejecta ($\sim 10^{51}-10^{52}$~erg), which expands outward at velocities of around $0.1 c$ \citep{1996snih.book.....A, branch2017supernova,2017AdAst2017E...5C}. The low-density outermost layer has the highest speed while the denser regions expand with slower velocities. After a few minutes, the SN ejecta reaches the companion NS, the hypercritical accretion starts. In the meanwhile, some matter falls back leading to an accretion process onto the $\nu$NS. This fallback accretion is significantly amplified by the companion NS which alters the trajectory of a partial SN ejecta that flows back to the $\nu$NS (Rueda et al., 2022 in preparation and \citealp{2019ApJ...871...14B}). The accretion rate onto the companion NS rises exponentially and peaks in a few minutes. The numerical simulations presented in \citet{2014ApJ...793L..36F,2016ApJ...833..107B,2019ApJ...871...14B} and (Rueda et al., 2022 in preparation) show that the entire hypercritical accretion process may last for hundreds of minutes, but the peak accretion rate of $\sim 10^{-3} M_\odot$ s$^{-1}$, supplied by the high density and slow-moving part of the SN ejecta, holds only for tens of seconds to tens of minutes depending on the binary separation, and the energy release is in the order of $10^{48}$--$10^{49}$~erg~s$^{-1}$. The accretion onto the $\nu$NS has two components, the first is the typical fallback matter analogous to the case of the SN of a single star, it leads the accretion rate to reach a peak to then decay nearly as a power-law with time, $\propto t^{-5/3}$. The peak luminosity produced by it is weak $<10^{48}$~erg~s$^{-1}$ and can be hardly for cosmological distances. The second is the unique feature of the binary system that is induced by the interaction of SN ejecta with the NS companion. The presence of the companion enhances the fallback onto the $\nu$NS creating a second peak of accretion \citep[see, e.g., Fig. 5 in]{2019ApJ...871...14B}. The second part contributes most to the accreting mass with an accretion rate of $\sim 10^{-3} M_\odot$ s$^{-1}$ at about an orbital-period time after the SN explosion (Rueda et al., 2022 in preparation). The fallback accretion also transfers angular momentum to the $\nu$NS spinning it up to a rotation period of a few milliseconds \citep{2017ApJ...835....4B}. The peak luminosity from the fallback accretion is in the order of $10^{48}$--$10^{49}$~erg s$^{-1}$, and occurs at minutes to tens of minutes after the SN explosion. As we show below, the fallback accretion will continue as a source of energy that powers the afterglow. The SN produces $\sim 0.4 M_\odot$ nickel whose radioactive decay energy is emitted mainly at optical wavelengths with a corresponding flux that peaks around $\sim 13$~day in the source rest-frame \citep{2017AdAst2017E...5C}. This optical signal can be observed from some low-redshift sources ($\sim z<1$) that are less affected by the absorption \citep{2006ARA&A..44..507W}. 

Summarizing, from the observational point of view, a few minutes after the SN explosion, we expect first to observe the signal from the accretion onto the companion NS and the $\nu$NS, their peak times may overlap or separate depending on the binary separation, hence there will be one or two pulses with the luminosities of the order of $10^{48}$~erg~s$^{-1}$ to $10^{49}$~erg~s$^{-1}$. Then, we will observe the afterglow emission due to the synchrotron emission from the SN ejecta with a luminosity that decays as a power-law, and at $\sim 13$~day the optical bump from the radioactive decay of nickel.

\section{Observation} 
\label{sec:observation}

At 19:55:53 UT, On 2019 August 29, GRB 190829A triggered the \textit{Fermi}-GBM \citep{2019GCN.25551....1F}. \textit{Swift}-BAT was triggered $51$~s later, but fortunately GRB 190829A was in the \textit{Swift}-BAT field of view before the trigger. Here in the paper, we take the GBM trigger time as $T_0$. The \textit{Swift}-XRT started to observe at time $T_0+148.3$~s \citep{2019GCN.25552....1D}. The redshift of $z=0.0785 \pm 0.005$ was proposed by \textit{Swift}-UVOT \citep{2019GCN.25552....1D}, Half Meter Telescope (HMT) \citep{2019GCN.25555....1X} and Nordic Optical Telescope (NOT) \citep{2019GCN.25563....1H} via associating to a nearby galaxy and later was confirmed by the spectroscopic observation of the GTC \citet{2021A&A...646A..50H}. GRB 190829A is one of the nearest GRBs ever observed. The SN association is found and confirmed by Liverpool Telescope and GTC telescope \citep{2019GCN.25657....1P, 2019GCN.25677....1D} and GROND \citep{2019GCN.25651....1B}.

We retrieve the \textit{Fermi} data from the Fermi Science Support Center\footnote{\url{https://fermi.gsfc.nasa.gov/ssc/}} (FSSC), and were analyzed using the Multi-Mission Maximum Likelihood framework (3ML)  \footnote{\url{https://threeml.readthedocs.io}} \citep{2015arXiv150708343V}. The spectrum fitting is performed by a Bayesian analysis with the Markov chain Monte Carlo (MCMC) iterations within the 3ML framework and the results are double-checked by implementing the Fermi GBM Data Tools \citep{GbmDataTools}. 
For a detailed Bayesian analysis of the data and the reduction procedure applied to a GRB spectrum, we refer to \citet{Li2019b,Li2019a,Li2019c,Li2021b,Li2021a}.
We retrieve the \textit{Swift} data from UK Swift Science Data Centre\footnote{\url{https://www.swift.ac.uk}} (UKSSDC), the analyzing and fitting are carried out by HEASoft\footnote{\url{https://heasarc.gsfc.nasa.gov/lheasoft/}} and 3ML. The VHE data observed by H.E.S.S. are from \cite{2020ApJ...898...42C}, the optical data observed by GTC are taken from \citet{2021A&A...646A..50H}, and the radio emission observed by Arcminute Microkelvin Imager - Large Array (AMI-LA) are taken from \citet{2020MNRAS.496.3326R}.

\section{Prompt Emission: SN explosion and NS accretion} 
\label{sec:prompt}

Both GBM and BAT light-curves show two pulses, see e.g. the \textit{Fermi}-GBM light-curve in Figs. \ref{fig:prompt_190829A} and \ref{fig:luminosity_190829A}. The first pulse starts to rise at time $-0.75$~s, peaks at $1.02$~s, and fades at $8.05$~s. The cutoff power-law function gives a best fit over the power-law and Band functions. We also tested the addition of a blackbody component to the above models but it did not lead to a statistical improvement of the fit. As shown in Fig. \ref{fig:prompt_spectrum}, the fit of the spectrum is characterized by the power-law index $\alpha = -1.15 \pm 0.06$ and peak energy $E_p = 144.28 \pm 50.67$~keV. The integrated isotropic energy\footnote{For the calculation of the luminosity distance, we use a Friedman-Lema\^itre-Robertson-Walker metric, Hubble constant $H_0=67.4\pm0.5$~km/s/Mpc, and matter density $\Omega_M = 0.315\pm0.007$ \citep{2018arXiv180706209P}.} from $1$~keV to $10$~MeV gives $4.25\pm1.02 \times 10^{49}$~erg~s$^{-1}$. The averaged luminosity is $4.84 \pm 1.16 \times 10^{48}$~erg~s$^{-1}$. 
After $38.45$~s, the second larger pulse rises at $46.50$~s, peaks at $51.65$~s and fades at $64.00$~s. This pulse is best fitted by a Band function with a low peak energy $E_p = 13.58 \pm 0.42$~keV that almost touches the lower edge of the \textit{Fermi}-GBM energy band. Because of the small amount of data of energy lower than $E_p$, the low-energy is unconstrained, in fact, we obtain $\alpha = 0.50 \pm 1.01$. The high-energy index $\beta = -2.53 \pm 0.02$ appears to be a typical value. The total energy in the second pulse is $3.56 \pm 0.50 \times 10^{50}$~erg, and the averaged luminosity is $2.05 \pm 0.29 \times 10^{49}$~erg~s$^{-1}$. Our spectral fit is consistent with the analysis of \citet{2021A&A...646A..50H}. 

We interpret these two pulses are due to the accretion onto the companion NS and the fallback accretion onto the $\nu$NS. The observed energy and luminosity are consistent with our expectation that the emission from the accretion processes with a luminosity of $\sim 10^{48}$--$\sim 10^{49}$~erg~s$^{-1}$. 

Numerical simulations of BdHNe show that the time evolution of the $\nu$NS fallback accretion rate has a two-peak structure, being the second peak a unique feature of the binary interactions while, the accretion onto the MNS companion, shows a single peak structure \citep[see, e.g., Fig. 5 in][]{2019ApJ...871...14B}. The first peak of the $\nu$NS fallback accretion is probably not observable because since before it occurs the star has little rotational energy to be released. Therefore, we assume that the two observed pulses are related to the second peak of the $\nu$NS accretion, and the peak of the MNS accretion. The simulations show that the fallback accretion rate onto the $\nu$NS weakly depends on the binary parameters, while the time of occurrence and intensity of the accretion peak onto the MNS depends crucially on the orbital period and the MNS initial angular momentum at the beginning of the accretion. The larger the orbital period the lower the MNS accretion peak and the later it occurs, approaching the time of occurrence of the second accretion peak of the $\nu$NS. The relatively short time separation between the two observed peaks in GRB 190829A suggests a binary period of the order of tens of minutes. This is also suggested by the energy released in the emission. For an orbital period in the range $20$--$40$ min, we expect a peak accretion rate of the MNS in the range $10^{-4}$--$10^{-5} M_\odot$ s$^{-1}$ \citep[see Fig. 5 in][]{2019ApJ...871...14B}, which translates into an accretion power of $10^{48}$--$10^{49}$ erg s$^{-1}$ assuming a $10\%$ of efficiency in the conversion from gravitational to radiation energy. If we assume the energy release is powered by the rotational energy gained during the accretion process, then we end with similar figures. The star gains angular momentum at a rate $\dot{J} \sim 2 \sqrt{3} G M \dot{M}/c \sim 4 \times 10^{45}$ g cm$^2$ s$^{-1}$, for a $1.5 M_\odot$ and the above accretion rate, which implies a spinup rate of about $40$ Hz min$^{-1}$. The simulations show that the MNS accretion peaks in about $1/10$ of the orbit, therefore for the above range of orbital periods, at the time of the accretion the MNS could rotate with a frequency of $80$--$160$ Hz, which implies a rotation power $\Omega \dot{J} \sim (2$--$5)\times 10^{48}$ erg s$^{-1}$, where $\Omega$ is the stellar angular velocity.

\section{Afterglow: Synchrotron and $\nu$NS pulsar radiation}\label{sec:synch}

Figure \ref{fig:luminosity_190829A} shows the multiwavelength luminosity light-curves. We notice the continuity of the \textit{Swift}-BAT and \textit{Swift}-XRT observations, and a soft X-ray depression at $\sim 10^2$--$10^3$~s after the prompt emission, then from $3 \times 10^5$~s the soft X-ray decays as a power-law of index $-1.26 \pm 0.06$. The optical and radio afterglows also have a power-law decay behaviour. The VHE evolves similarly to the X-ray, with a luminosity of $\sim 25\%$ of the absorption-corrected X-ray luminosity, similarly to other GRBs with VHE observations \citep{2019Natur.575..464A,2019Natur.575..455M,2019Natur.575..448Z,abdalla2021revealing}. 

The optical observations show an additional bump after $10^6$~s \citep{2019GCN.25657....1P, 2019GCN.25677....1D}, which indicates the SN optical emission powered by nickel decay. 

We here follow and extend the treatment of the GRB afterglow by \citet{2018ApJ...869..101R} within the BdHN scenario. In this picture, the afterglow originates from the synchrotron radiation produced by the expansion of the SN ejecta in presence of the magnetic field of the $\nu$NS. We now estimate the emission generated by the synchrotron mechanism in the X-rays, in the optical, and in the radio, together with the pulsar emission of the $\nu$NS. The $\nu$NS contributes the energy of afterglow by two means, first the fallback accretion of surrounding matters, the energy from which dominates the early afterglow, and second the release of rotation energy of its spin-down, which produces the late-time X-ray afterglow. This model predicts that the VHE emission is not directly emitted by this synchrotron emission, although it seems to be related to the $\nu$NS activity (see Sec. \ref{sec:conclusion}). 

\subsection{Synchrotron emission by the expanding ejecta}\label{sec:synch1}

Because the electrons lose their energy very efficiently by synchrotron radiation, we can apply a one-zone model assuming that the radiation originates from the ejecta, say $r=R_*$. We assume the ejecta expands at constant velocity $v_{*,0}$, so the radius evolves as
\begin{equation}\label{eq:radius}
    R_*(t) = R_{*,0}\,\hat{t},
\end{equation}
where $\hat{t} \equiv t/t_*$, and $t_* \equiv R_{*,0}/v_{*,0}$.

In agreement with pulsar theory \citep[see, e.g.,][]{1969ApJ...157..869G, 1969ApJ...157.1395O}, we assume that at large distances from the $\nu$NS, beyond its light-cylinder, the magnetic field decreases linearly with distance. This implies that the magnetic field strength felt by the expanding ejecta evolves with time as
\begin{equation}\label{eq:B}
    B_*(t) = B_{*,0}\, \frac{R_{*,0}}{R_*} = \frac{B_{*,0}}{\hat{t}},
\end{equation}
where $B^{(0)}_*$ is the magnetic field strength at $r=R_{*,0}$, and we have used Eq. (\ref{eq:radius}).
 
The evolution of the distribution of radiating electrons per unit energy, $N(E,t)$, is obtained from the solution of the well-known classical kinetic equation \citep[see, e.g.][]{1962SvA.....6..317K}
\begin{equation}\label{eq:kinetic}
    \frac{\partial N(E, t)}{\partial t}=-\frac{\partial}{\partial E}\left[\dot{E}\,N(E,t)\right] + Q(E,t),
\end{equation}
that accounts for the particle energy losses, and $Q(E,t)$ is the number of injected electrons per unit time, per unit energy, and $\dot E$ is the electron energy loss rate. In the present case, the electrons are subjected to the adiabatic losses due to the ejecta expansion and to synchrotron radiation losses because of the magnetic field. Therefore, the electron energy evolves with time according to the classical energy balance equation \citep{1962SvA.....6..317K}
\begin{equation}\label{eq:gammadot}
    -\dot E = \frac{E}{t} + P_{\rm syn}(E,t),
\end{equation}
where
\begin{equation}\label{eq:Psyn}
    P_{\rm syn}(E,t) = \beta B_*^2(t) E^2,
\end{equation}
is the bolometric synchrotron power, and $\beta = 2e^4/(3 m_e^4 c^7)$  \citep[see, for details, e.g.,][]{2011hea..book.....L}.

We adopt a distribution of the injected particles following a power-law behavior \citep[see, e.g.,][]{1962SvA.....6..317K, 1979rpa..book.....R, 2011hea..book.....L}
\begin{equation}\label{eq:Q}
Q(E,t)=Q_0(t)E^{-\gamma}, \qquad 0\leq E \leq E_{\rm max},
\end{equation}
where $\gamma$ and $E_{\rm max}$ are parameters to be determined from the observational data. We now address the function determining the rate of particle injection, $Q_0(t)$, which is related to the power injected by the $\nu$NS into the ejecta, i.e. the injected electrons are accelerated by the energy from the fallback accretion onto the $\nu$NS. We assume that the bolometric power released by the $\nu$NS is given by
\begin{equation}\label{eq:Lt}
L_{\rm inj}(t) = 
L_0 \left(1+\frac{t}{t_q}\right)^{-k},
\end{equation}
where $L_0$, $t_q$, and $k$ are model parameters. Because the ejecta surrounds the $\nu$NS, the power released by the $\nu$NS is injected into the ejecta, so the function $Q_0(t)$ can be found from energy conservation as
\begin{align}\label{eq:LandQ}
L_{\rm inj}(t) &= \int_{0}^{E_{\rm max}} E\,Q(E,t) dE =Q_0(t)\frac{E_\mathrm{max}^{2-\gamma}}{2-\gamma},
\end{align}
that via Eq.~(\ref{eq:Lt}) leads to
\begin{equation}\label{eq:Q0}
    Q_0(t) =
q_0\left(1+\frac{t}{t_q}\right)^{-k},
\end{equation}
where $q_0 \equiv  (2-\gamma)L_0/E_{\rm max}^{2-\gamma}$.

With the specification of the ejecta evolution given by Eq.~(\ref{eq:radius}), of the magnetic field by Eq.~(\ref{eq:B}), and the rate of particle injection given by Eqs.~(\ref{eq:Q}) and (\ref{eq:Q0}), we proceed to integrate the kinetic equation (\ref{eq:kinetic}). For this task, we must first find the time evolution of the energy of a generic electron injected at time $t=t_i$ with initial energy $E_i$. With all the above, Eq.~(\ref{eq:gammadot}) is a Riccati's differential equation that has the following analytic solution \citep{2022arXiv220200314R, 2022arXiv220200316R}
\begin{equation}\label{eq:gammavst}
    E = \frac{E_i\,(t_i/t)}{1 + {\cal M} E_i t_i\left( \frac{1}{\hat{t}_i^{2}} -  \frac{1}{\hat{t}^{2}}\right)},
\end{equation}
where ${\cal M}\equiv \beta B^2_{*,0}/2$.

Following \citet{1973ApJ...186..249P}, we write the solution of Eq. (\ref{eq:kinetic}) as
\begin{equation}\label{eq:Nsol}
    N(E,t) = \int_E^\infty Q[E_i, t_i(t,E_i,E)] \frac{\partial t_i}{\partial E} dE_i,
\end{equation}
where $t_i(t,E_i,E)$ is obtained from Eq. (\ref{eq:gammavst}). The solution $N(E,t)$ can be written as a piecewise function of time depending upon the behavior of the energy injection (\ref{eq:Q0}), i.e. at times $t<t_q$ it can be approximated as a constant, while at longer times it is well approximated by a pure power-law function. In addition, as we shall show below, the GRB afterglow data is well explained by a regime in which synchrotron radiation losses dominate over adiabatic losses. Under these conditions, the solution of Eq. (\ref{eq:Nsol}) can be written as \citep{2022arXiv220200314R}
\begin{align}\label{eq:N3}
&N(E,t)\approx \begin{cases}
    \frac{q_0}{\beta B_{*,0}^2 (\gamma-1)}\hat{t}^{2} E^{-(\gamma+1)}, & t < t_q\\
   \frac{q_0 (t_q/t_*)^{k}}{\beta B_{*,0}^2 (\gamma-1)}\hat{t}^{2-k} E^{-(\gamma+1)}, &   t_q < t < t_b,
\end{cases}
\end{align}
where the electron's energy is in the range $E_b < E < E_{\rm max}$, being 
\begin{equation}\label{eq:Eb}
    E_b = \frac{\hat{t}}{{\cal M} t_*},\quad t_b = t_*^2 {\cal M}  E_{\rm max}.
\end{equation}

The synchrotron luminosity radiated in the frequencies $[\nu_1,\nu_2]$ can be then obtained as
\begin{equation}\label{eq:Lnu}
    L_{\rm syn}(\nu_1,\nu_2; t) = \int_{\nu_1}^{\nu_2} J_{\rm syn}(\nu,t)d\nu,
\end{equation}
where $J_{\rm syn}(\nu,t)$ is the synchrotron spectral density (energy per unit time, per unit frequency), $\nu_1=\nu$, $\nu_2=\nu+\Delta\nu$, being $\Delta\nu$ the bandwidth. The synchrotron power is mostly emitted at radiation frequencies close to the so-called critical frequency $\nu_{\rm crit} = \alpha B_* E^2$, where $\alpha = 3 e/(4\pi m_e^3 c^5)$ \citep[see, e.g.,][]{1979rpa..book.....R}. Therefore, the bolometric synchrotron power (\ref{eq:Psyn}) can be readily written in terms of the radiation frequency $\nu$ as
\begin{equation}\label{eq:Psynbol}
    P_{\rm syn}(E,t) \approx P_{\rm syn}(\nu,t) = \frac{\beta}{\alpha} B_* \nu = \frac{\beta}{\alpha} \frac{B_{*,0}}{\hat{t}}\nu,
\end{equation}
and within the same approximation \citep[see, e.g.,][]{2011hea..book.....L} 
\begin{equation}
    J_{\rm syn}d\nu\approx P_{\rm syn}(\nu,t) N(E,t)dE.
\end{equation}
We now replace this into Eq. (\ref{eq:Lnu}) and obtain the synchrotron luminosity
\begin{equation}\label{eq:Lnu2}
    L_{\rm syn}(\nu, t) = \frac{\beta}{2} \alpha^{\frac{p-3}{2}} \eta B_{*,0}^{\frac{p+1}{2}}\hat{t}^{\frac{2 l-p-1)}{2}}\nu^{\frac{3-p}{2}},
\end{equation}
where we have used the approximation $\Delta\nu/\nu\ll 1$ in the integration of Eq. (\ref{eq:Lnu}) in view of the power-law behavior of $J_{\rm syn}$, and we have written the electron distribution as $N(E,t) = \eta\,\hat{t}^l E^{-p}$, being $\eta$, $l$ and $p$, known constants from Eq. (\ref{eq:N3}).

\begin{figure*}
\centering
\includegraphics[width=0.75\hsize,clip]{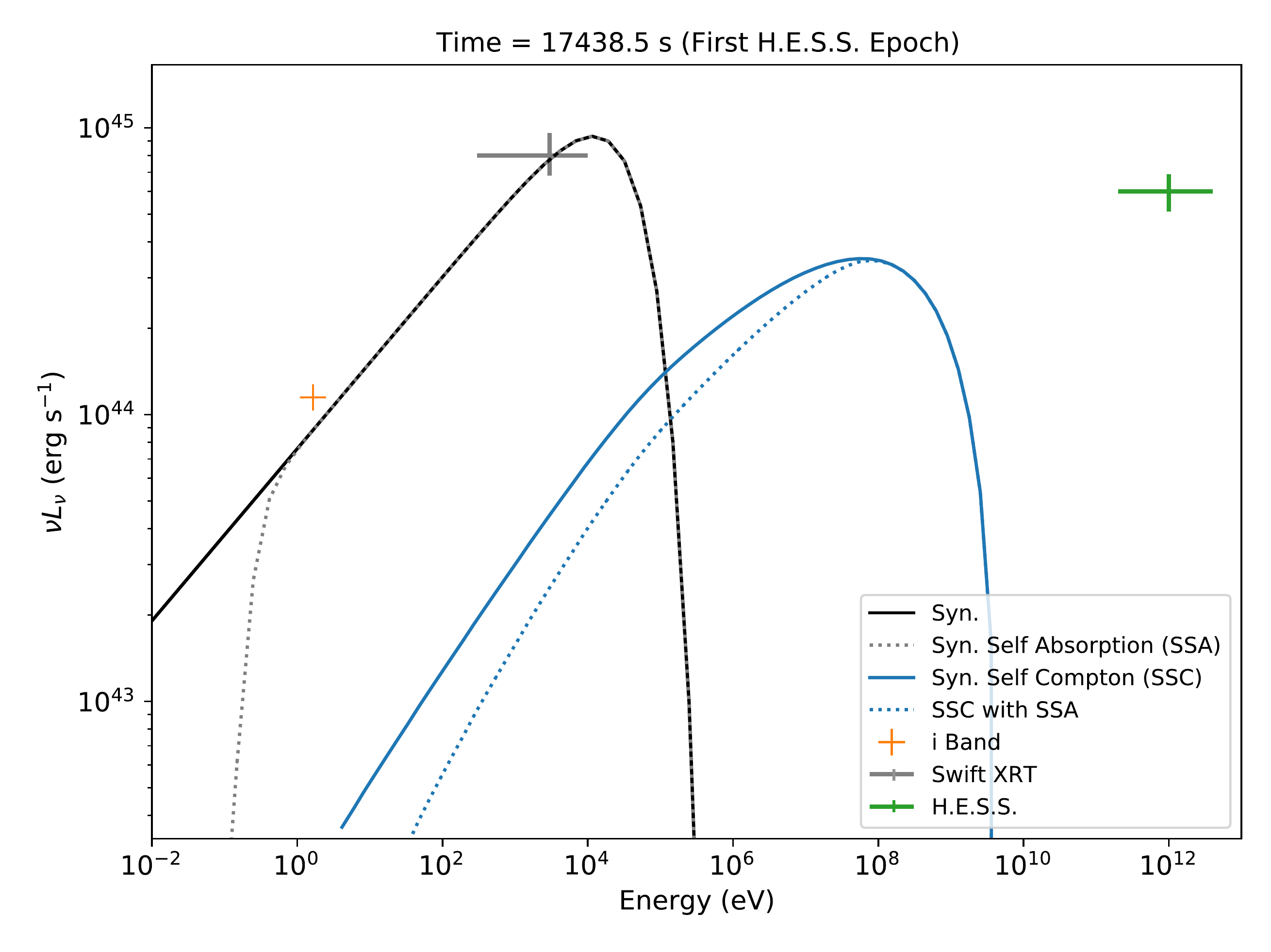}
\caption{The synchrotron and SSC spectra of the first H.E.E.S observational epoch. The synchrotron well fits the optical and X-ray data. Neither the synchrotron nor the SSC is able to produce the H.E.S.S. observations.}
\label{fig:SSC}
\end{figure*}

Therefore, the synchrotron power has a power-law dependence both in time and radiation frequency, see Eq. (\ref{eq:Lnu2}). If the system remains over time in the same physical regime in which the energy losses of the electrons are dominated by synchrotron radiation, the luminosity in the X-rays, optical, and radio wavelengths decrease with the \textit{same} power-law index (see Fig. \ref{fig:fit190829A}).  For the parameters of GRB 190829A (see Table \ref{tab:parameters}), we found that this condition is fulfilled, i.e., the afterglow data remains at times $t<t_b$, and is explained by electron energies that hold in the range $E_b<E<E_{\rm max}$, see Eq. (\ref{eq:Eb}). In this case, the ratio of the synchrotron luminosity at different frequencies is constant in time because depends only on the power-law index of the injection rate as \citep{2022arXiv220200316R}
\begin{equation}\label{eq:Lratio}
    \frac{L_{\rm syn} (\nu_1,t)}{L_{\rm syn} (\nu_2,t)} = \left( \frac{\nu_1}{\nu_2} \right)^{\frac{3-p}{2}} = \left( \frac{\nu_1}{\nu_2} \right)^{\frac{2-\gamma}{2}}.
\end{equation}

In practice, we fix the value of $\gamma$ from the X-rays to optical luminosity ratio. Having fixed $\gamma$, the optical (or X-rays) to radio luminosity ratio is fixed too. Figure \ref{fig:fit190829A} shows that this procedure leads to a synchrotron luminosity in the radio band that also agrees with the observations. This result implies that this model correctly described the  afterglow in the wide range of energies, in addition to the X-ray, also the radio and the optical, giving a strong support to the proposed scenario for the afterglow emission.

\subsection{$\nu$NS evolution and pulsar emission}\label{sec:synch2}

\begin{figure*}
\centering
\includegraphics[width=0.75\hsize,clip]{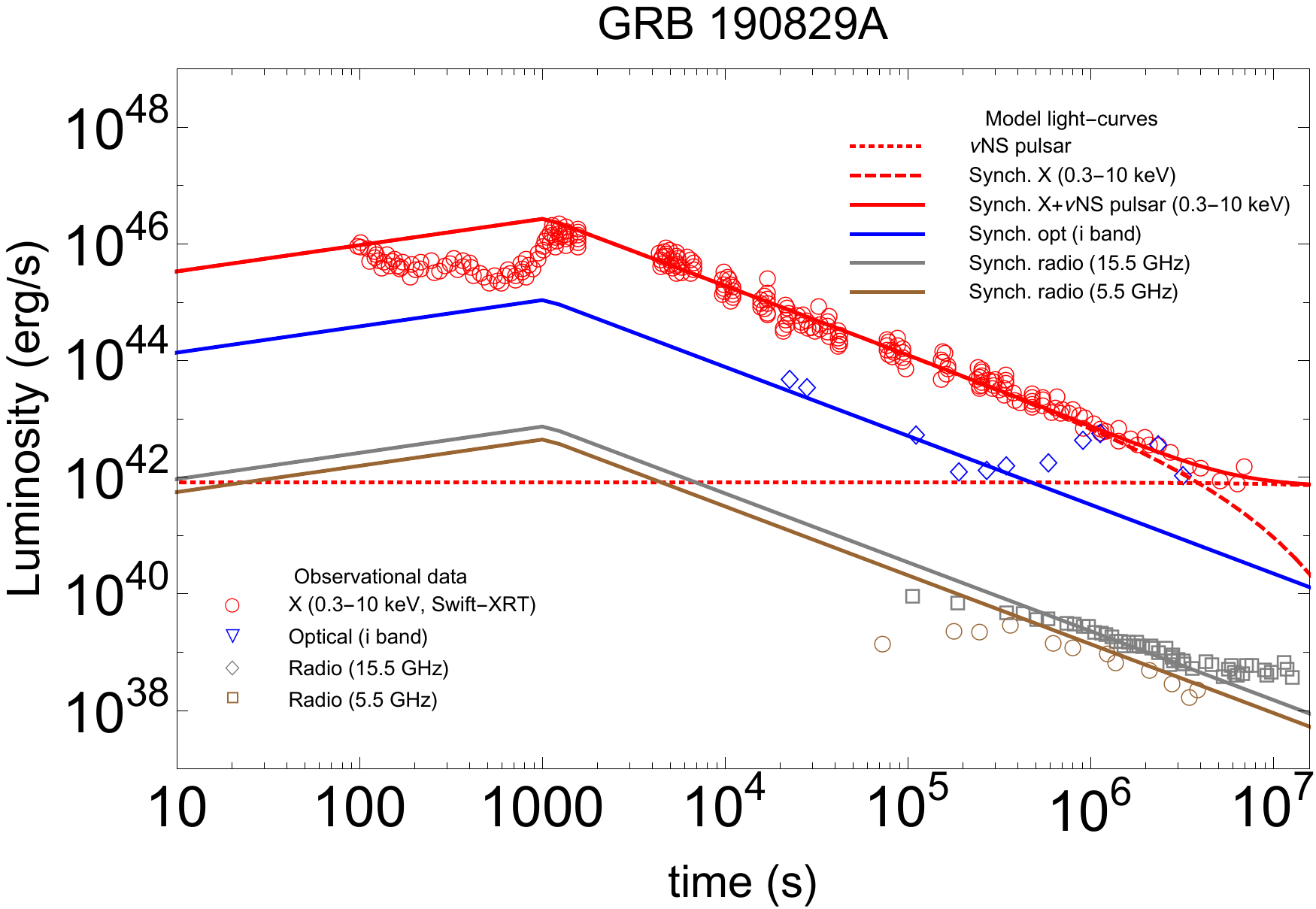}
\caption{Luminosity of GRB 190829A in the X-rays ($0.3$--$10$ keV), optical (i band) \citep{2021A&A...646A..50H}, and in the radio ($5.5$ and $15.5$ GHz) energy bands \citep{2020MNRAS.496.3326R}.}
\label{fig:fit190829A}
\end{figure*}

As the synchrotron luminosity fades with time, the pulsar-like emission of the $\nu$NS and the MNS companion becomes observable in the X-ray afterglow. We expect the magnetic field of the younger $\nu$NS dominates over that of the much older MNS companion. By the time of the BdHN event, the MNS magnetic field could have decayed with respect to its birth value. Although microphysical mechanisms leading to magnetic field decay in pulsars have been debated, a relevant mechanism for such a decay is that during the evolution, the binary passes through common-envelope and X-ray binary phases in which the magnetic field is reduced by long-term accretion episodes \citep[see, e.g.][and references therein, for numerical simulations]{2007MNRAS.376..609P}.

Bearing the above in mind, we assume the pulsar emission observable in the afterglow is driven by the magnetic field of the $\nu$NS. We calculate this pulsar emission following the dipole+quadrupole magnetic field model presented in \citet{2015MNRAS.450..714P}. The total pulsar (spindown) luminosity is obtained by summing the dipole and quadrupole contributions
\begin{align}\label{eq:Lsd}
    L_{\rm sd} &= L_{\rm dip} + L_{\rm quad} \nonumber \\
    &= \frac{2}{3 c^3} \Omega^4 B_{\rm dip}^2 R^6 \sin^2\chi_1 \left( 1 + \xi^2 \frac{16}{45} \frac{R^2 \Omega^2}{c^2} \right),
\end{align}
with $R$ the $\nu$NS radius, and $\xi$ defines the quadrupole to dipole strength ratio
\begin{equation}
    \xi \equiv \sqrt{\cos^2\chi_2+10\sin^2\chi_2} \frac{B_{\rm quad}}{B_{\rm dip}},
\label{eq:eta}
\end{equation} 
where the modes can be separated as: $\chi_1 = 0$ and any value of $\chi_2$ for the $m = 0$ mode, $(\chi_1, \chi_2) = (90^\circ, 0^\circ)$ for the $m = 1$ mode, and $(\chi_1, \chi_2) = (90^\circ, 90^\circ)$ for the $m = 2$ mode. 

The evolution of the $\nu$NS is calculated from by integrating the energy balance equation
\begin{equation}\label{eq:Erot}
	-(\dot{W}+\dot{T}) = L_{\rm tot} = L_{\rm inj} + L_{\rm sd},
\end{equation}
where $W$ and $T$ are, respectively, the $\nu$NS gravitational and rotational energy. 

\begin{table}
    \centering
    \begin{tabular}{l|r}
    Parameter & Value \\
    \hline
      $\gamma$ &  $1.01$\\
       $k$  & $1.63$\\
       $L_0$ ($10^{46}$ erg s$^{-1}$)& $8.00$\\
       $E_{\rm max}$ ($10^4 \ m_e c^2$) & $5.00$\\
       $t_q$ (s) & $1050.00$\\
       $R_{*,0}$ ($10^{11}$ cm) & $1.00$ \\
       $v_{*,0}$ ($10^{9}$ cm s$^{-1}$) & $1.00$ \\
       $B_{*,0}$ ($10^{6}$ G) & $5.01$\\
       $\xi$ & $100.00$\\
       $B_{\rm dip}$ ($10^{12}$ G) & $5.00$ \\
       $P$ (ms) & $8.00$\\
       \hline
    \end{tabular}
    \caption{Numerical values of the theoretical model of synchrotron radiation that fit the multiwavelength observational data of GRB 190829A as shown in Fig. \ref{fig:fit190829A}.}
    \label{tab:parameters}
\end{table}

Table \ref{tab:parameters} lists the model parameters that fit the afterglow of GRB 190829A in the X-rays, optical \citep{2021A&A...646A..50H}, and radio energy bands \citep{2020MNRAS.496.3326R} as shown in Fig. \ref{fig:fit190829A}. The power-law luminosity observed in the multiwavelength data after $10^3$ s is well explained by the synchrotron emission. We do not find evidence up to times $10^7$ s of a change in the power-law index, which implies that the system up to these times has not transitioned yet to the physical regime of the dominance of adiabatic losses over synchrotron losses. Although there is a good agreement of the model with the data after $10^3$ s, the fit of the emission $<10^3$ s is complicated. At those times, the behavior of the light-curve is far from smooth, which is likely due to factors other than the synchrotron alone. The modeling of this early part of the afterglow is challenging for the present simplified synchrotron picture, and probably needs very detailed information on the density profile in the ejecta and absorption processes that we are not considering. The light-curve at early times may catch short timescale details of the evolution, so the accurate evaluation of the absorption and/or scattering processes (e.g. synchrotron self-absorption or Thomson scattering) leading to the evolution of the optical depth might need detailed radiative transfer calculation including possible deviations of the density and expansion velocity from spherical symmetry (e.g. polar angle dependence and filaments arising from Rayleigh-Taylor instability), and/or of the thermodynamics variables (e.g. temperature and composition) needed for the evaluation of the opacity at every photon energy, position, and time. In addition, the early evolution of the $\nu$NS could be highly complex leading to an energy injection that deviates from the law assumed in Eq. (\ref{eq:Lt}). The latter implies a constant injection rate at times $t \lesssim \tau_q \approx 10^3$ s (see Table \ref{tab:parameters}) leading to the rising synchrotron luminosity following a power-law at those times (see Fig. \ref{fig:fit190829A}).

The VHE emission observed in the $0.2$--$4$~TeV energy band of H.E.S.S. is not explained by the above synchrotron model. We now estimate whether the synchrotron self-Compton radiation (SSC) could originate such an emission. The SSC emission is produced by synchrotron photons that upscatter off the relativistic electrons that produce them. The upscattering increases the energy of those photons by a factor equal to the square of the electron Lorentz factor, leading to a spectrum with a shape similar to the synchrotron spectrum but at higher energies \citep{2009herb.book.....D,zhang2018physics,2019ApJ...884..117W,2022A&A...660A..18N}. Figure \ref{fig:SSC} shows as an example the first observational epoch of H.E.S.S. ($17438.5\pm805.5$~s) and our estimate the SSC emission for the parameters of our synchrotron model. The SSC emission peaks at a few hundreds of MeV, cutoffs at $<10$~GeV, and has a lower luminosity with respect to the observed in the H.E.S.S. energy bandwidth. Therefore, we conclude that neither the synchrotron nor the SSC explain the VHE emission of GRB 190829A observed by H.E.S.S. However, the similar power-law behavior of the VHE and the X-ray light-curves suggests the former could be related to some (at the moment unexplored) transient activity of the $\nu$NS. We notice that the H.E.S.S. team expressed a similar conclusion that the traditional afterglow model including SSC does not explain their observations, and they expected a multi-zone emission model \citep{abdalla2021revealing}.

We turn now to the synchrotron emission. The critical synchrotron radiation energy ($h \nu_{\rm crit}$) decreases with time, so the peak of the synchrotron radiation shifts to lower energies with time. Around $10^6$ s, the critical radiation energy falls below the keV range leading to the exponential decay of the synchrotron emission in the X-rays after that time. Subsequently, the pulsar emission from the $\nu$NS dominates the observed X-ray emission. We have taken advantage from this behavior to infer the strength of the dipole and quadrupole components of the magnetic field, as well as the rotation period of the $\nu$NS. 

The bump observed in the optical data at about $10^6$ s is explained by the SN emission powered by the energy released from nickel decay \citep{1996snih.book.....A}, which in this specific GRB, the type Ic SN 2019oyw optical signal overshoots the synchrotron optical emission. We refer the detailed SN observation and analysis to the article from GTC \citep{2021A&A...646A..50H}.

The radio emission shows some excess over the synchrotron emission from a few $10^6$ s to $10^7$ s. This feature may be a signature from the $\nu$NS pulsar, although further observational data and theoretical analysis are needed to confirm this hypothesis.

\section{Discussion and Conclusion}
\label{sec:conclusion}

The BdHN scenario describes the late evolution of a CO$_{\rm core}$-NS binary. In particular, it predicts the electromagnetic signals that can be observed from a sequence of episodes triggered when the CO$_{\rm core}$ undergoes gravitational collapse at the end of its thermonuclear evolution, generating a SN and forming a $\nu$NS at its center. The SN ejected material accretes onto the companion NS and also onto the $\nu$NS via matter fallback. The fate of the companion NS depends on the initial mass and crucially on the binary separation (i.e., on the orbital period) that sets the accretion rate. BdHN I are characterized by short orbital periods of the order of a few minutes, where the NS reaches by accretion the critical mass for gravitational collapse into a BH. We refer the reader to \citet{2021MNRAS.504.5301R} for a comprehensive analysis of $380$ BdHN I. In this article, we have analyzed GRB 190829A, which is classified as a BdHN II. These sources are characterized by longer orbital periods, i.e., larger binary separations, with lower accretion rates and therefore the companion NS does not reach the critical mass for gravitational collapse.

GRB 190829A, at the close distance of redshift $0.0785$, was observed by multi-band telescopes and satellites on the ground and in space. These detailed observations have given us the opportunity to find the emissions that correspond to the episodes expected to occur in a BdHN II. The initial X-ray pulse of energy $\sim 4.25 \times 10^{49}$~erg and the second pulse of energy $\sim 3.56 \times 10^{50}$~erg represent the SN ejecta accretion onto the companion NS and the $\nu$NS, see figures \ref{fig:prompt_190829A} and \ref{fig:prompt_spectrum} for their light-curves and spectra.

We explained the radio, optical and X-ray afterglow emissions as due to the synchrotron radiation from the SN ejecta expanding in the magnetic field of $\nu$NS. The $\nu$NS continuously inject energy into the SN ejecta from fallback accretion and the spin-down owing to magnetic braking. From the fitting of afterglow synchrotron emission, see figure \ref{fig:fit190829A},  we infer the $\nu$NS spinning at a $8$~ms period with a dipole field of $5\times10^{12}$~G. The VHE emission observed is not explained either by this synchrotron radiation process or by the SSC. However, the fact that the VHE light-curve shows a similar power-law decay to the X-rays, with a lower luminosity released but at higher photon energy, is suggestive of a process related to a transient activity of the $\nu$NS, e.g. \textit{glitches}, that shares a portion of the rotational energy and lead to a narrow-angle emission near the light-cylinder. The modeling of such complex physical phenomenon needs further theoretical work and simulations, and as such, goes beyond the scope of the present article. This same VHE emission parallel to the X-ray afterglow has been observed as well in GRB 180720B \citep{2021arXiv210309158M} and GRB 190114C (Ruffini, et al., to be submitted).

The BdHN model naturally contains an SN, and indeed in GRB 190829A it was observed the SN association. The peak of the SN standard optical luminosity (Moradi, et al., to be submitted) is higher than the synchrotron optical emission, see figure \ref{fig:luminosity_190829A}, which makes the optical SN signal distinguishable.

In general, this article presents the evolutionary picture of the late stage of a binary system, which produces a GRB induced by a SN. We have observed two pulses of luminosity $\sim 10^{49}$~erg~s$^{-1}$ from the accretion from the SN ejecta onto the NS and the $\nu$NS, as well as the NS spin-down. From the observations, we infer the $\nu$NS has an initial spin of $8$~ms and dipole magnetic field $5\times10^{12}$~G.


\acknowledgements
Y.A. acknowledges funding from the Science Committee of the Ministry of Education and Science of the Republic of Kazakhstan (Grant No. AP08855631).

\bibliographystyle{aasjournal}
\bibliography{190829A}

\end{document}